\documentclass[prl,aps,showpacs,showkeys,groupedaddress,superscriptaddress,twocolumn]{revtex4-2}

\usepackage[utf8x]{inputenc}
\usepackage{color}
\usepackage{bbm} 
\usepackage{amsfonts,amsmath,amssymb,stmaryrd}
\usepackage{graphicx}
\usepackage{subfigure}  % use for side-by-side figures
\usepackage{hyperref}
\usepackage{epsfig}
\usepackage{mathrsfs}
\usepackage{verbatim}
\usepackage{centernot}

\renewcommand{\a}{\hat{a}}

\newcommand{\ad}{\hat{a}^\dagger}

\usepackage{array}

\usepackage{bm}	% bold in math mode

\begin{document}

%%%%%%%%%%%%%%%%%%%%%%%%%%%%%%%%%%%%%%%%%%%%%%%%%
\title{Finite-temperature topological invariant for interacting systems}
%%%%%%%%%%%%%%%%%%%%%%%%%%%%%%%%%%%%%%%%%%%%%%%%%

\author{Razmik Unanyan}
\affiliation{Department of Physics and Research Center OPTIMAS, University of Kaiserslautern, 67663 Kaiserslautern, Germany}
\author{Maximilian Kiefer-Emmanouilidis}
\affiliation{Department of Physics and Research Center OPTIMAS, University of Kaiserslautern, 67663 Kaiserslautern, Germany}
\affiliation{Department of Physics, University of Manitoba, Winnipeg, Manitoba, Canada R3T 2N2}
\author{Michael Fleischhauer}
\affiliation{Department of Physics and Research Center OPTIMAS, University of Kaiserslautern, 67663 Kaiserslautern, Germany}
\affiliation{Kavli Institute for Theoretical Physics, University of California, Santa Barbara, California 93106, USA}

%%%%%%%%%%%%%%%%%%%%%%%%%%%%%%%%%%%%%%%%%%%%%%%%%

\begin{abstract}
We generalize the Ensemble Geometric Phase (EGP), recently introduced to classify the 
topology of density matrices, 
to finite-temperature states of \emph{interacting} systems in one spatial dimension (1D).
This includes cases where the gapped ground state has a fractional filling and is degenerate.
At zero temperature the corresponding topological invariant agrees with the well-known invariant of
Niu, Thouless and Wu.
We show that its value at finite temperatures is identical to that of the ground state below some critical temperature $T_c$ larger than the many-body gap.
We illustrate our result with numerical
simulations of the 1D extended super-lattice Bose-Hubbard model at quarter filling. Here a cyclic change
of parameters in the ground state leads to a topological charge pump with fractional winding $\nu=1/2$. The particle transport is no longer quantized 
when the temperature becomes comparable to the many-body gap, yet the winding of the generalized EGP is. 
\end{abstract}


\date{\today}
\maketitle

%%%%%%%%%%%%%%%%%%%%%%%%%%%%%%%%%%%%%%%%%%%%%%%%%

%%%%%%%%%%%%%%%%%%%%%%%%%%%
\paragraph{Introduction. --}
%%%%%%%%%%%%%%%%%%%%%%%%%%%

Starting with the discovery of the quantum Hall effect \cite{Klitzing-PRL-1980,TKNN-PRL-1982,Tsui-PRL-1982,Laughlin-PRL-1983,Arovas-PRL-1984}
topology has become an important paradigm for the understanding and classification of phases of matter \cite{Xiao-RMP-2010,Hazan-Kane-RMP-2010,Wen-RMP-2017}. 
Topology is characterized by integer-valued invariants which describe global properties of the system and are responsible
for the robustness of characteristic features like quantized bulk transport, or edge states and edge currents. 
Invariants such as the winding of the geometric 
%Berry or 
Zak phase or the Chern number of single-particle Bloch functions
are defined in terms of the wave-function of ground states and are thus restricted to cases where the system 
is in a 
%can be described by a 
pure state.

In recent years several attempts have been made to generalize the concept of topology to finite-temperatures and to non-equilibrium 
steady states of non-interacting fermions \cite{Avron-NJP-2011,Bardyn-NJP-2013,Viyuela-PRL-2014,Huang-PRL-2014,Viyuela-PRL-2014b,Nieuwenburg-PRB-2014,Linzner-PRB-2016,Grusdt-PRB-2017,Bardyn-PRX-2018,Altland2020}. For example, it was shown that the generalization of geometric
phases to density matrices based on the Uhlmann construction \cite{Uhlmann-Rep-Math-Phys-1986} leads to consistent topological
invariants in 1D \cite{Viyuela-PRL-2014,Huang-PRL-2014}. Its application to higher dimensions \cite{Viyuela-PRL-2014b} is however faced with
difficulties \cite{Budich-Diehl-PRB-2015}. Other approaches predict an unphysical extensive number of topological phases for arbitrarily small temperatures
\cite{Grusdt-PRB-2017}.

Recently it was shown that the winding of the many-body polarization introduced by Resta \cite{Resta-PRL-1998} is an alternative
 topological invariant for Gaussian mixed states of fermions in 1D, termed \emph{ensemble geometric phase} (EGP) \cite{Linzner-PRB-2016,Bardyn-PRX-2018,Bardyn-2017}, which can also be
applied in 2D \cite{Wawer}.
It was shown that the EGP of finite-temperature states in non-interacting systems is reduced to the ground-state 
Zak phase in the thermodynamic limit $L\to\infty$ and thus these states have the same topological 
classification  as the corresponding ground states (following the Altland-Zirnbauer classification \cite{Altland-PRB-1997,Schnyder-PRB-2008,Ryu-NJPhys-2010}). Despite being a genuine many-body quantity the EGP can be measured directly \cite{Bardyn-PRX-2018}. Furthermore a
non-trivial winding of the EGP of a finite-temperature or non-equilibrium steady state upon cyclic parameter variations has direct 
physical consequences. E.g. it can lead to quantized transport in a weakly coupled auxiliary system initially prepared in a low temperature state \cite{Wawer-2}. For non-interacting bosons the EGP winding is always zero \cite{Mink-PRB-2019}.

In the present paper we extend this concept to the case of interactions between particles including the possibility of
interaction-induced fractionalization and degeneracy. We show that a generalization of the EGP
to systems with a gapped ground state of fractional filling allows to define a topological invariant for
finite-temperature states of one-dimensional systems of \emph{interacting} particles. The winding of the EGP reduces to the well-known
Niu-Thouless-Wu (NTW) invariant \cite{Niu-JPhysA-1984,Niu-PRB-1985} at $T= 0$ and has the same value for all temperatures below a certain critical value. 
% We argue that the critical temperature is at least on the order of the many-body gap. 
The EGP also provides a theoretical tool to 
detect topological order present in the ground state
in cases where the gap is small  or 
numerical calculations are restricted to non-zero temperatures.
We illustrate our results with numerical
simulations of the extended super-lattice Bose-Hubbard model (Ext-SLBHM)  \cite{ExtSLBHM} at quarter filling, where the ground state is a doubly degenerate Mott insulator (MI).  

%%%%%%%%%%%%%%%%%%%%%%%%%%%%%%
\paragraph{Ensemble Geometric Phase: integer case. --}
%%%%%%%%%%%%%%%%%%%%%%%%%%%%%%

We consider  1D lattice models with a many-body hamiltonian $H(\lambda)$ that has a periodic dependence on some variable $\lambda$. We 
assume periodic boundary conditions in space with $L$ unit cells and allow for interactions among the particles. Since single-particle 
Bloch states $\vert u_k\rangle$ are no longer a good eigen-basis, topological invariants must be defined
in terms of the many-body ground state $\vert \Phi_0\rangle$.  As suggested by Niu, Thouless and Wu \cite{Niu-PRB-1985},  the Zak phase of Bloch wavefunctions
can be generalized by replacing the single-particle crystal momentum $k$ 
by a twist angle $\theta$ 
of boundary conditions $\Phi(x_1,\dots x_j + L ,\dots x_N) = e^{i \theta}\Phi(x_1,\dots  x_N)$. Since $\theta$ and $\theta+2\pi$ define the same boundary conditions, the parameter space $(\theta,\lambda)$ is a torus ${\sf T}^2$. Twisted boundary conditions can be removed and replaced by periodic ones via a canonical transformation to a twisted Hamiltonian, $\bar H(\theta) = \hat U(\theta) H \hat U(\theta)^{-1}$ and $\vert \Psi_0(\theta)\rangle = \hat U(\theta) \vert \Phi_0(\theta)\rangle$. 
Here
\begin{equation}
\hat U(\theta) = e^{ i \theta \hat X},\quad \textrm{with}\quad\hat X = \frac{1}{L}\sum_{j=1}^L\sum_{s=1}^n \left(j + r_s\right) \hat n_{js}
\end{equation}
is the momentum shift operator with $\hat n_{js}$ denoting the particle number operator of the $s$th site $(s\in\{1,2,\dots,n\})$ in the $j$th unit cell, and the
lattice constant is $a=1$. $0\le r_s \le 1$ characterizes the position within the unit cell.
In terms of the $\vert\Psi\rangle$'s the many-body equivalent of the Zak phase then reads
$\phi_\textrm{MB} = i \int_0^{2\pi}\!\! \! d\theta \, \, \bigl\langle \Psi_0(\theta)\bigr\vert \partial_\theta \Psi_0(\theta)\bigr\rangle.$

If $\vert \Psi_0(\theta)\rangle$ is a gapped and \emph{non-degenerate} ground state, a slow (adiabatic) change of the parameter $\lambda=\lambda(t)$ in a closed loop, such that the many-body gap does not close, induces a Thouless pump, described by a current density
$\langle \hat j\rangle  = \partial_\theta E(\theta) / \hbar + i \bigl(\langle \partial_t \Psi_0\vert \partial_\theta \Psi_0\rangle -\langle \partial_\theta\Psi_0\vert \partial_t \Psi_0\rangle\bigr).$
Then following Niu, Thouless and Wu, averaging over twisted boundary conditions one finds a strictly integer-quantized particle transport 
over one time period ${\cal T}$. The transported charge $\Delta n$ is then directly
related to the NTW invariant $\nu$
\begin{eqnarray}
\Delta n  &=& \frac{1}{4\pi} \int_0^{\cal T}\!\!\! dt \int_0^{2\pi}\!\!\! d\theta\,   i \left[\Bigl\langle \partial_t \Psi_0\Bigr\vert \partial_\theta \Psi_0\Bigr\rangle -
\Bigl\langle \partial_\theta\Psi_0\Bigr\vert \partial_t \Psi_0\Bigr\rangle\right]\nonumber\\
&=& \frac{1}{2\pi} \int_0^{\cal T}\!\!\! dt \, \frac{\partial \phi_\textrm{MB}}{\partial t}=\frac{1}{2\pi} \oint d\lambda \, \frac{\partial \phi_\textrm{MB}}{\partial \lambda}=
\nu,\label{eq:NTW}
\end{eqnarray}
which is the generalization of the celebrated invariant of Thouless, Kohmoto,  Nightingale and den Nijs (TKNN) for free fermions \cite{Thouless-PRB-1983,TKNN-PRL-1982} 
to the case of interactions and disorder.
In Ref.\cite{Bardyn-PRX-2018} the TKNN invariant was generalized to Gaussian mixed states of fermions using 
the King-Smith -- Vanderbildt relation \cite{King-Smith-PRB-1993} between
changes of $\phi_\textrm{MB}$ to those of the many-body polarization 
$ P = \frac{1}{2\pi} \textrm{Im} \log \bigl\langle\Psi_0\bigr\vert \hat U\bigl\vert \Psi_0\bigr\rangle$
introduced by Resta \cite{Resta-PRL-1998}, where  $\hat U\equiv \hat U(2\pi)$:
$\partial_\lambda \phi_\textrm{MB}=2 \pi \partial_\lambda P$. 
This then allowed to replace the ground-state average  by the trace
over a density matrix
 $\phi_\textrm{EGP} = 2\pi P = \textrm{Im} \log \textrm{Tr}\bigl\{ \rho \, \hat U\bigr\}$ defining the so-called \emph{ensemble geometric phase} in \cite{Bardyn-PRX-2018}.

%%%%%%%%%%%%%%%%%%%%%%%%%%%%%%
\paragraph{Ensemble Geometric Phase: fractional case.--}
%%%%%%%%%%%%%%%%%%%%%%%%%%%%%%

In the absence of interactions, gapped ground states (of fermions) occur only at integer fillings per unit cell. This changes with
interactions. Here gapped ground states can exist which have fractional fillings and the Lieb-Schulz-Mattis theorem and its generalizations \cite{Lieb-AnnPhys-1961,Oshikawa-PRL-2000}
tell us that they attain a "topological order" accompanied by fractionalization and \emph{degeneracy}
\cite{Tao-PRB-1984}. 
In such a case the above arguments do not hold as a parameter loop
of $\lambda$ does not return the initial state to itself (up to a phase), but in general to an orthogonal state in the ground-state manifold. 
For a $d$-fold degenerate subspace the NTW invariant (\ref{eq:NTW}) must instead be replaced by the gauge-invariant determinant of a
Wilson loop \cite{Wilczek-PRL-1984}
\begin{equation}
\nu_\textrm{tot} = \frac{1}{2\pi} \int_0^{\cal T}\!\!\! dt\, \frac{\partial}{\partial t} \textrm{Im}\log \det {\sf W}(t),\quad {\sf W} ={\cal P} e^{i\int_0^{2\pi}\!\! d\theta\,  {\sf A}(\theta)}.
\nonumber
%\label{eq:nu_tot}
\end{equation}
Here ${\sf A}_{\mu\nu}(\theta) = i\bigl\langle \Psi_0^\mu\bigr\vert\partial_\theta\Psi_0^\nu\bigr\rangle$ is a $d\times d$ matrix,
and ${\cal P}$ denotes path ordering.  

We now argue that also $\nu_\textrm{tot}$ can be related to an expectation value of a unitary operator, which then allows for 
a generalization to mixed states. To see this, we note that following Niu, Thouless and Wu \cite{Niu-PRB-1985}, $\nu_\textrm{tot}$ can be expressed
as integral of the Berry curvature corresponding to any one of the degenerate ground states $\vert\Psi_0^\mu\rangle$ over an enlarged torus, extending either the
time integration to ${\cal T} \cdot d$ or the $\theta$ integration to $2\pi \cdot d$, ($\mu=0,\dots,d-1$)
\begin{equation}
\nu_\textrm{tot} = \frac{1}{2\pi} \int_0^{{\cal T}} \!\!\! dt   \int^{2\pi d}_0\!\!\! d\theta \, \, \textrm{Im} \, 
\langle \partial_t \Psi_0^\mu \vert \partial_\theta \Psi_0^\mu\rangle.
\label{eq:nu-tot}
\end{equation}
As a consequence particle transport 
is integer quantized only after $d$ cycles of a Thouless pump \cite{Thouless-PRB-1983}. 

As shown by Aligia and Ortiz \cite{Aligia-PRL-1999,Aligia-EPL-1999,Ortiz-pss-2000}, eq.(\ref{eq:nu-tot}) gives also the winding number of 
a many-body Berry phase
\begin{equation}
\nu_\textrm{tot} =\oint \frac{d\lambda}{2\pi} \frac{\partial \phi^{(d)}_\mu}{\partial \lambda},\quad\textrm{with}\enspace
\phi^{(d)}_\mu =  \textrm{Im}\log \bigl\langle\Psi_0^\mu\bigr\vert ({\hat U})^d \bigl\vert \Psi_0^\mu\bigr\rangle,\label{eq:EGP-degenerate}
\end{equation}
which does not depend on the particular ground state. To see this, we note that the lattice Hamiltonian is invariant under spatial translation
by one unit cell, described by the unitary lattice shift operator $\hat T$. For a $d$-fold ground-state degeneracy one can construct a basis set of ground states $\{\vert \Phi_0^\mu\rangle\}$ with $\vert \Phi_0^\mu\rangle = \hat T^\mu \vert \Phi_0^0\rangle$ and $\mu =0,\dots, d-1$. Since $\hat T^{-1} \hat U^d\hat T = \hat U^d e^{2\pi i d N/L}$ with $N$ being
the total number of fermions, and $d N/L$ is an integer for a fractional filling $1/d$, the Berry phases $\phi^{(d)}_\mu$  of all states in this basis have the same value. 
%For the same reason it is not surprising that  expression (\ref{eq:EGP-degenerate}) contains $(\hat U)^d$  rather than $\hat U$.
It should be noted that the finiteness of the absolute value $\vert \langle \Psi\vert (\hat U)^d\vert\Psi\rangle\vert $ in the thermodynamic limit 
is an indicator of a localized, i.e. insulating ground state  with filling $1/d$ per unit cell \cite{Resta-Sorella-PRL-1999,Aligia-PRL-1999}.

Eq.(\ref{eq:EGP-degenerate}) then allows to define a generalized ensemble geometric phase for mixed states
\begin{equation}
\phi_\textrm{EGP}^{(d)} =  \textrm{Im} \log \textrm{Tr}\bigl\{ \rho \, (\hat U)^d\bigr\}\label{eq:EGP-general}
\end{equation}
which we now use to construct a topological winding number for mixed states.
% of interacting one-dimensional systems. 
% Note that since the Berry phase for every  ground state in the basis $\{\vert \Phi_0^\mu\rangle\}$ is the same,
% eq.\eqref{eq:EGP-general} gives the same value for the EGP, $\phi_\textrm{EGP}^{(d)}$, of the degenerate ground state $\rho_0 = \frac{1}{d}\sum_\mu \vert \Phi_0^\mu\rangle\langle\Phi_0^\mu\vert $ as eq.\eqref{eq:EGP-degenerate}.

%%%%%%%%%%%%%%%%%%%%%%%%%%%%%%
\paragraph{Extended super-lattice Bose-Hubbard model. --}
%%%%%%%%%%%%%%%%%%%%%%%%%%%%%%

\begin{figure}[h]
\includegraphics[width=\linewidth]{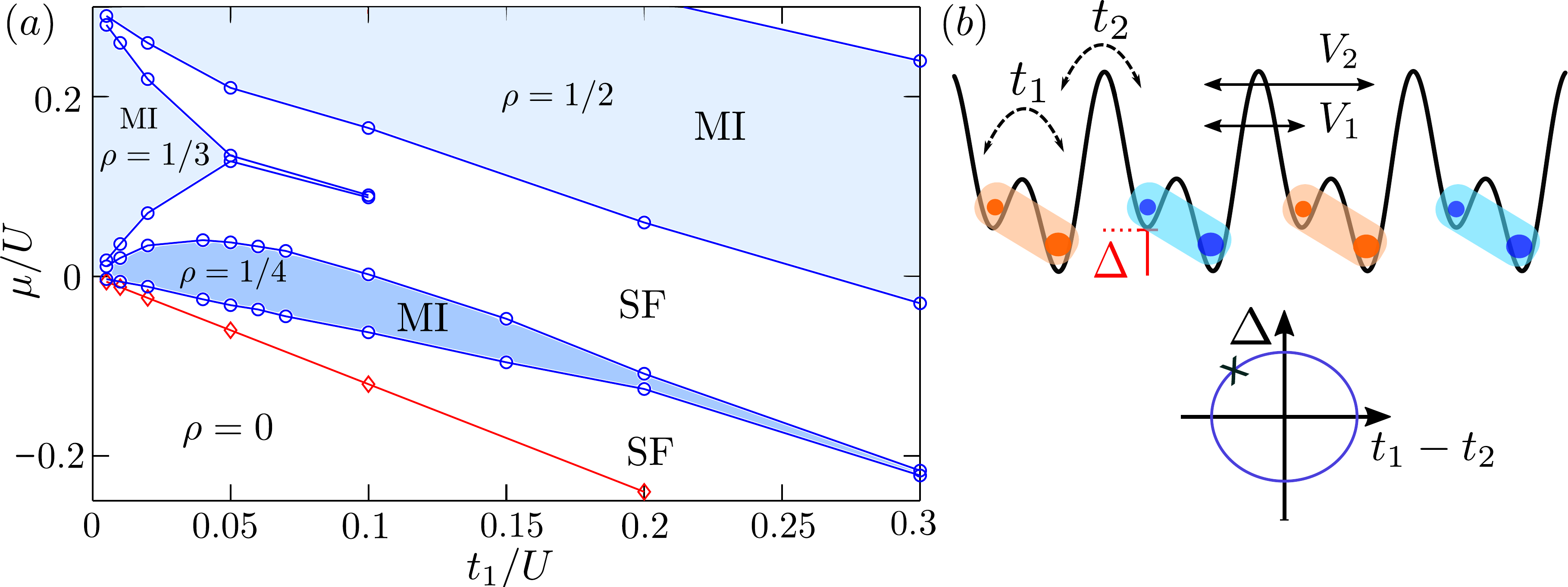}
\caption{(Color online) (a) Phase diagram of Ex-SLBHM for $t_2=0.5 t_1$, $V_1= 2 V_2 = 0.2 U$. Blue and white areas indicate Mott-insulating (MI) and superfluid (SF) phases, respectively.
The MI at average filling $\rho = 1/4$ per lattice site is doubly degenerate 
corresponding to superpositions of two density waves indicated in (b) in blue and orange. (b) Cyclic adiabatic variations of $t_1 - t_2$ and $\Delta$ encircling the
point $\Delta = t_1-t_2=0$ lead to a fractionally quantized charge pump \cite{Zheng-PRB-2016,Li-PRB-2017}
}
\label{fig:system}
\end{figure}

Let us discuss a specific example with interaction-induced fractional topological charges and
associated fractional winding number \cite{Zheng-PRB-2016,Li-PRB-2017}
, the one-dimensional extended super-lattice Bose Hubbard model (Ext-SLBHM).  As shown in
Fig.\ref{fig:system}(b), bosons at lattice site $j$, described by annihilation and creation operators $\a_j$ and $\ad_j$, and with onsite interaction strength $U$ move along a 1D lattice with alternating hopping  $t_1$ and $t_2$ and a staggered potential $\Delta$. In addition there is a nearest-neighbor (NN) and next-nearest neighbor (NNN)
interaction $V_1$ and $V_2$ respectively. The Hamiltonian reads in second quantization
\begin{eqnarray}
&& H = -t_1 \sum_{j,\textrm{even}} \ad_j \a_{j+1} - t_2 \sum_{j,\textrm{odd}} \ad_j \a_{j+1} + h.a. \\
  && + \frac{\Delta}{2} \sum_j (-1)^j \ad_j\a_j+ \frac{U}{2}\sum_j 
 \hat n_j(\hat n_j-1) + \sum_{j,d} V_d \, \hat n_i \hat n_{i+d}.\nonumber
 \end{eqnarray}
Here $\hat n_j = \ad_j \a_j$. The system has a unit cell of two sites. With periodic boundary conditions and in the absence of interactions one finds two single-particle energy bands with a finite gap, which closes only for $\Delta=t_1-t_2=0$. 
For sufficiently strong interactions there are further gap openings and Mott insulating (MI) states with fractional fillings 
% per unit cell 
emerge. The 
ground-state phase diagram, obtained from DMRG simulations \cite{DMRG} is
shown in Fig.\ref{fig:system}(a), where MI phases with fractional fillings are indicated.

%%%%%%%%%%%%%%%%%%%%%%%%%%%%%%%%%%
%%%%%%%%%%%%%%%%%%%%%%%%%%%%%%%%%%%

\begin{figure}[h]
\includegraphics[width=\linewidth]{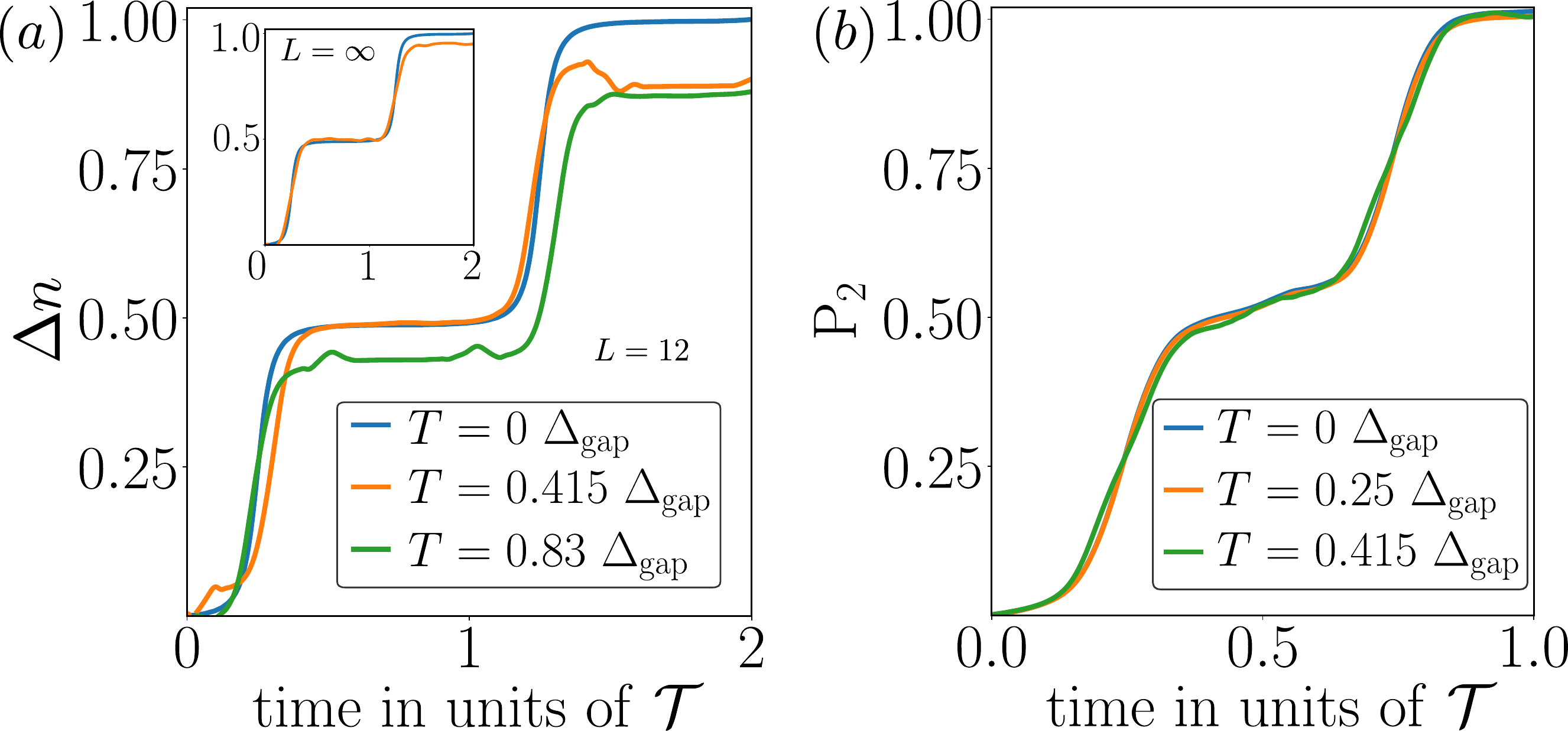}
\caption{(Color online) (a) Integrated particle current in the Ext-SLBHM as function of time
for a small system of $L =12$ unit cells and different temperatures obtained by exact diagonalization in the $\rho=1/4$ MI phase. 
Here $t_{1/2}=5(1\pm\cos\lambda(t)$, and $\Delta = -60 \sin\lambda(t)$, with angle $\lambda(t) =2\pi t/{\cal T}+3\pi/2$ and 
$V_1=40$, $V_2=20$, and $U$ is infinite.
The insert shows the same for $L\to\infty$ obtained by LCRG (Light-Cone Renormalization Group, see supplementary material \cite{supplement} for details) \cite{LCRG1,LCRG2,LCRG3}.
(b) Generalized EGP for the two-fold degenerate system also obtained by LCRG. One notices that the winding remains strictly quantized even at temperatures where there is a substantial occupation of excited states.}
\label{fig:winding}
\end{figure}

In the following we are interested in the phase with average
filling of $\rho=1/4$ per lattice site, or $1/2$ per unit cell. Here the ground state is doubly degenerate for periodic boundary conditions if the number $L$ of unit cells is even. If one starts in one of the two many-body ground states, say $\vert\Psi_1\rangle$, and changes  $t_1-t_2$ and  $\Delta$ in a closed loop in parameter space such that the many-body gap remains open at all points, the
ground-state manifold $\{\vert \Psi_1\rangle, \vert \Psi_2\rangle\}$ returns to itself up to a $U(2)$ rotation. 
Then $2$  loops in parameter space need to be performed for the initial state to return to itself modulo a phase. Similarly the average current $\langle \hat j \rangle$ needs to be integrated over $2$ periods of length ${\cal T}$
to lead to an integer-quantized number $\Delta n$ of pumped particles, which is verified by our numerical simulations
in Fig.\ref{fig:winding}a. 
This does no longer hold true, however, for finite temperatures. As expected and shown in the same figure the number of transported particles deviates substantially from unity as soon as the temperature approaches the many-body gap, since higher energy states are occupied.

Remarkably and in sharp contrast, the winding of the generalized EGP $\phi^{(2)}_\textrm{EGP}$ remains strictly unity even at temperatures on the order of the many-body gap as can be seen from Fig.\ref{fig:winding}b.  (It should be noted that increasing the temperature becomes numerically more demanding.)

%%%%%%%%%%%%%%%%%%%%%%%%%%%%%%%%%
%%%%%%%%%%%%%%%%%%%%%%%%%%%%%%%%%%%

It was shown in Ref.\cite{Bardyn-PRX-2018} for non-interacting fermions, that the winding of the EGP $\phi_\textrm{EGP}$ remains the same for all temperatures $T<\infty$. In the following we will give some arguments that also for interacting systems the EGP winding of a thermal state is identical to that of the ground state below some critical temperature which is larger than the many-body gap.

%%%%%%%%%%%%%%%%%%%%%%%%%%%%%%
\paragraph{Finite-temperature winding. --}
%%%%%%%%%%%%%%%%%%%%%%%%%%%%%%

We first show that there exists a critical temperature $T_c$ different from zero, below which the winding of the 
EGP coincides with that of the many-body ground state $\Delta \phi_\textrm{EGP}\vert_{T<T_c} = \Delta \phi_\textrm{EGP}\vert_{T=0}$.
Thus, different e.g. from the prediction in \cite{Grusdt-PRB-2017}, a temperature-induced topological transition can
only occur at a finite, nonzero temperature.

To this end we note that the change of $\phi_{\textrm{EGP}}$
\begin{equation}
\Delta\phi_{\textrm{EGP}}=2\pi\deg z=\operatorname{Im}{\displaystyle\oint\limits}
\frac{1}{z\left(  \lambda\right)  }\frac{dz\left( \lambda \right)  }{d\lambda}
d\lambda \label{Polarization_change}
\end{equation}
with $z(\lambda)\ne 0$ for $\lambda\in \delta{\cal C}$,
is just the winding number or degree of a smooth map $z(\lambda)=$Tr$\{\rho(\lambda)\,
(\hat U)^{d}\}$ from a closed loop $\delta {\cal C}$ in parameter space $(\lambda)$ 
into the complex plane $\mathbb{C}$ of the polarization amplitude $z(\lambda)$. It
measures the algebraic change of phase of $z$ as the variable $\lambda$ goes around
the loop once. The degree of $z$ is by definition the number of
solutions of $z=0$ inside $\delta {\cal C}$ taking into account their algebraic
multiplicity (for more details about degree theory see \cite{Regan}). Suppose
for all temperatures $T$ with $T_1< T < T_2$, $z\neq0$ on $\delta{\cal C}$, then 
$\Delta\phi_{\textrm{EGP}}$ is independent of temperature in the interval $(T_1,T_2)$. 
%Indeed, it is clear from eq. (\ref{Polarization_change}) that $\Delta\phi_{\textrm{EGP}}$ is a continuous function of $T$. 
These arguments are in parallel with Hopf's homotopy theorem
\cite{Regan}. Now, the condition $z(T,\lambda)\neq0$ everywhere on $\delta{\cal C}$ 
can always be fulfilled in a finite range of temperatures starting at $T=0$ until for a critical temperature $T_c$ 
\begin{equation}
\bigl\vert z(T_c,\lambda)\bigr\vert =0, \label{eq:Tc}
\end{equation}
for some $\lambda$ on $\delta{\cal C}$. This defines a \emph{temperature-induced topological phase transition}.

%%%%%%%%%%%%%%%%%%%%%%%%%%%%%%
\paragraph{Critical temperature. --}
%%%%%%%%%%%%%%%%%%%%%%%%%%%%%%

\begin{figure}[t]
\includegraphics[width=0.98\linewidth]{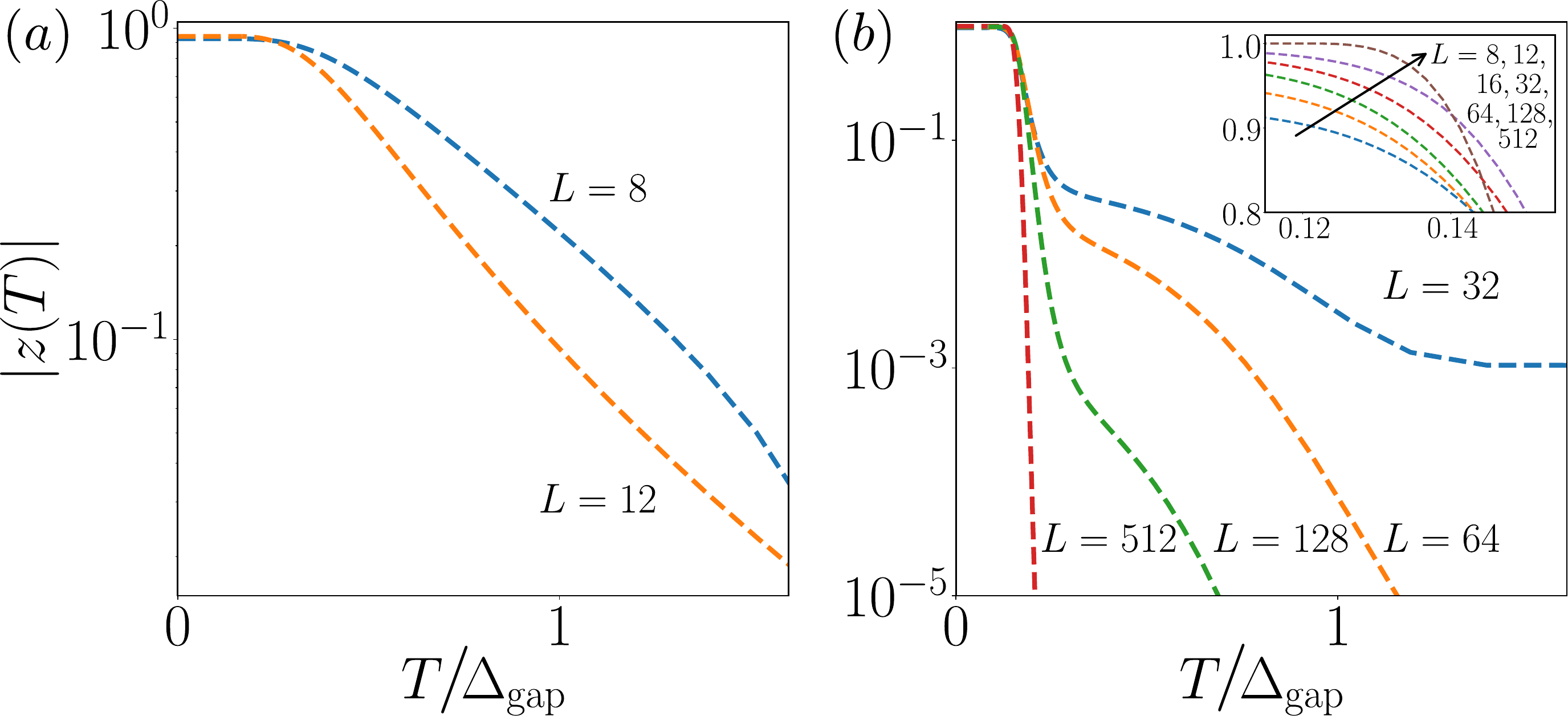}
\caption{(Color online) Temperature scaling of polarization amplitude $\vert z(T)\vert$ for the extended SLBHM for the parameters of fig.\ref{fig:winding}
and different system sizes
(a) for small systems obtained with exact diagonalization, (b) for larger systems obtained with LCRG, which is an infinite size method.
Here $\lambda(t)=3\pi/4$ but we verified that the results hold for any value of $\lambda$.
Also a finite length $L$ was cut out and used for the calculation of $\vert z(T)\vert$. The inset shows the behaviour in the low temperature regime.
As shown in \cite{Resta-Sorella-PRL-1999} $\vert z(T\to 0)\vert \to 1$ for increasing system size.
}
\label{fig:scaling}
\end{figure}

As shown in \cite{Bardyn-PRX-2018}, $T_c=\infty$ for non-interacting fermions. To investigate the possibility of a finite-temperature
topological phase transition in an interacting system, we have numerically calculated $\vert z(T)\vert $ for the extended SLBHM. The results are shown
in Fig.\ref{fig:scaling} in dependence of $T$ for different system sizes. One recognizes that $\vert z\vert $ remains 
approximately unity below temperatures that are a sizable fraction of the many-body gap  $\Delta_\textrm{gap}$ followed by a fall-off, which
becomes more pronounced with increasing system size. This behavior is very similar to the non-interacting case, for which one finds a double-exponential
scaling (see supplementary material) $\vert z(T)\vert  \sim \exp\bigl(-2L\exp(-\beta \Delta_\textrm{gap})\bigr)$. Note, that although $z(T)\to 0$ for  $L\to \infty$, 
the temperature $T_0$, where $\vert z\vert$ starts to deviate from unity only scales logarithmically 
with system size $L$.
\begin{equation}
T_0 \,\sim \, {\Delta_\textrm{gap}}/{\ln L}.
\end{equation} 
From our numerical data it remains unclear if $\vert z(T)\vert$ has a strict zero, indicating a topological phase transition, or not
and it would be interesting to investigate other interacting superlattice models such as \cite{Torio-PRB-2006,Stenzel-PRA-2019}.

In order to show that interactions may lead to a finite value of $T_c$ let us consider the \textit{flattened} Hamiltonian 
\begin{equation}
\tilde H = E_0 \sum_{\mu=1}^d \vert \Psi_0^\mu\rangle \langle \Psi_0^\mu\vert + E_1 \sum_{j\ne (0,\mu)} \vert \Psi_j\rangle \langle \Psi_j\vert,
\end{equation}
with  $\Delta_\textrm{gap} = E_1-E_0$. Here all excited states have the maximum weight compatible with temperature and gap. 
Since the total Chern number of all excited bands must be opposite to that of the ground state, flattening a many-body Hamiltonian 
to such a form  is expected to lead to the
"worst" case.
Then for the polarization amplitude $z(T)=\textrm{Tr}\{\exp(-\beta \tilde H) (\hat U)^d\}/Z$ holds
$\textrm{Tr}\{e^{-\beta \tilde H} (\hat U)^d\} =  z(0) d \left(e^{-\beta E_0} -e^{-\beta E_1}\right) 
+ \textrm{Tr}\{ (\hat U)^d\} e^{-\beta E_1} $,
where $z(0) =  \frac{1}{d}\sum_\mu \langle \Psi_0^\mu\vert (\hat T)^d\vert \Psi_0^\mu\rangle $ is the zero-temperature polarization amplitude.
As shown in \cite{Resta-Sorella-PRL-1999,Aligia-PRL-1999} $\vert z(0)\vert \to 1$ in the thermodynamic limit $L\to\infty$ if the ground state is insulating.
Since $\textrm{Tr}\{(\hat U)^d\}$ does not depend on the Hamiltonian, its phase is fixed and does not change upon parameter variations. 
Thus the winding of $\phi_\textrm{EGP}^{(d)}$ for the flattened Hamiltonian $\tilde H$ at temperature $T$ remains equal to that in the ground state as long as
$e^{-\beta E_0}-e^{-\beta E_1} > \vert \textrm{Tr}\{(\hat U)^d\}\vert e^{-\beta E_1}/d$. With this we find for the critical temperature $T_c$, defined
through eq.\eqref{eq:Tc}, (with $k_B=1$)
\begin{equation}
T_c = \frac{\Delta_\textrm{gap}}{\ln\left(1+|\textrm{Tr}\{(\hat U)^d\}| /d\right)}.
\end{equation}
As shown in the supplementary material $\vert \textrm{Tr}\{(\hat U)^d\}\vert$ is intensive and bounded by a value of order ${\cal O}(d)$.

%%%%%%%%%%%%%%%%%%%%%%%%%%%%%%%%%%
%%%%%%%%%%%%%%%%%%%%%%%%%%%%%%%%%%%

%%%%%%%%%%%%%%%%%%%%%%%%%%%%%%%%%%
\paragraph{Summary. --}
%%%%%%%%%%%%%%%%%%%%%%%%%%%%%%%%%%

We have introduced a many-body topological invariant for finite-temperature states of interacting many-body systems
with fractional filling and ground-state degeneracy. The invariant is based on a generalization of the EGP introduced in \cite{Bardyn-PRX-2018}. 
In the limit $T\to 0$ it coincides with the well-known NTW invariant \cite{Niu-PRB-1985}. We showed that there exists a critical temperature $T_c$, defined by a
vanishing polarization amplitude, which generically is larger than the many-body gap. Below $T_c$ the EGP-based topological winding number is identical to the ground-state invariant.
% There are some indications that the critical temperature is infinite in the generic case. 
The generalized EGP also allows to detect topological properties from finite-temperature measurements. It  
can be measured either directly (see \cite{Bardyn-PRX-2018}) or obtained from the full counting statistics, measurable e.g. in ultra-cold atom experiments with a gas microscope.
A non-trivial topological winding of the EGP can furthermore induce a quantized transport in a weakly coupled auxiliary system, prepared at a low temperature \cite{Wawer}.
We illustrated our results for the Ext-SLBHM model at quarter filling, which has a two-fold degenerate ground state
and an associated fractional topological charge of $1/2$. The arguments given here can be extended to interacting two-dimensional systems of the Chern class
with translational invariance, and thus to systems like fractional Chern insulators with intrinsic topological order. These systems can be mapped to independent one-dimensional systems 
by transforming to momentum space in one of the spatial directions. 

%%%%%%%%%%%%%%%%%%%%%%%%%%%%%%%%%%
\subsection*{acknowledgement}
Financial support from the DFG through SFB TR 185, project number 277625399  is gratefully acknowledged. 
The authors thank Sebastian Diehl and Matthias Moos for stimulating discussions and Richard Jen for help in the DMRG calculation of the ground-state
phase diagram. We also thank J. Sirker for support with the LCRG code. M.F. thanks the KITP for hospitality, where
part of this work was done and supported in part by the National Science Foundation under Grant No. NSF PHY-1748958.

%%%%%%%%%%%%%%%%%%%%%%%%%%%%%%%%%%%%%%%%%%%%%%%%%%%%%%%%%%%%%%%%%
\newpage
\onecolumngrid
\section{Appendix}

%%%%%%%%%%%%%%%%%%%%%%%%%%%%%%%%%%%%%%

\section{Numerical methods}

%%%%%%%%%%%%%%%%%%%%%%%%%%%%%%%%%%%%%%

\begin{figure}[h]
	\includegraphics[width=0.8\textwidth]{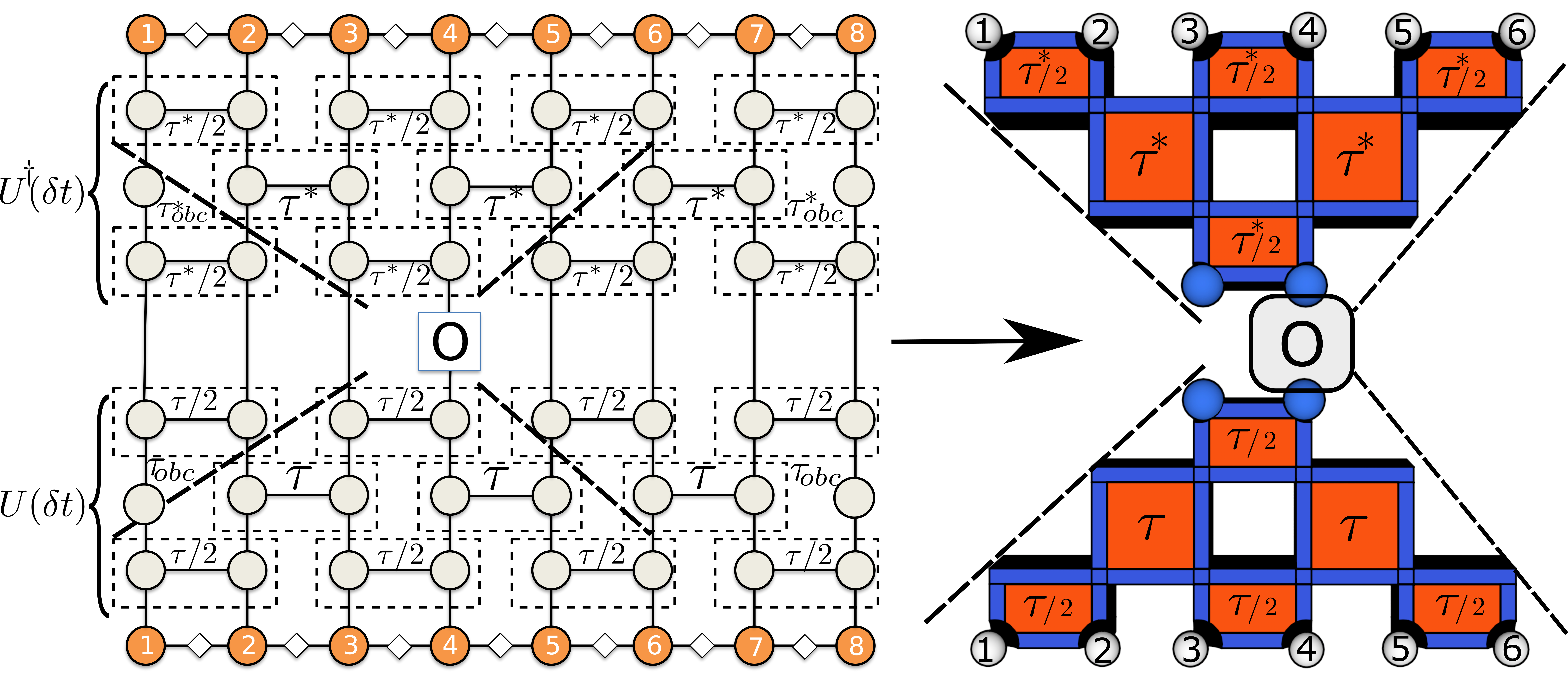}
	\caption{A time evolved MPS network in Vidal form (left) is transformed to a light cone shape (right). The time evolution follows a second order trotter-Suzuki decomposition. Here adjacent plaquetts outside of the light cone cancel each other. This picture has been taken and been modified from Fig. 2 in \cite{LCRG3}.}
	\label{MPStolightcone}
\end{figure}

For computations beyond exact diagonalization routines we used the light cone renormalization group (LCRG) \cite{LCRG2}, which allows to calculate observables directly in the thermodynamic limit. To this end the LCRG   restricts the time-evolution of an operator to the Hilbert space of an effective light cone using the Lieb-Robinson bound for a Hamiltonian with short range interactions \cite{lieb1972}, see Fig. \ref{MPStolightcone}. Besides applying the Lieb-Robinson bounds, the Hilbert space is truncated similarly to time-adaptive density matrix renormalization group (t-)DMRG algorithms, in its traditional formulation by S.R. White \cite{white1992,white2004}. The formalism can however also be understood  in the language of matrix product states (MPS) \cite{DMRG}. To  time-evolve the system stepwise the LCRG algorithm uses transfer matrices to enlarge the system size and a second order Trotter-Suzuki decomposition \cite{Trotter,Suzuki1}. Connecting to conventional MPS language, a tensor network can be easily transformed to the  light cone shape as adjacent plaquettes would just trace out, see Fig \ref{MPStolightcone}. A description of the full algorithm can be found in Ref. \cite{LCRG2} as well as a working code in the supplementary material of Ref. \cite{LCRG2}. 
The algorithm has been generalized to open systems in Refs. \cite{LCRG3,LCRG4}. In particular temperature effects are included by extending the Hilbert space by auxiliary degrees of freedom and subsequent purification. This is explained in detail in Refs. \cite{LCRG1,LCRG5}.\\ 

In this supplementary we will not go into further  details of the LCRG time-evolution, for which we refer to \cite{LCRG2,LCRG3,LCRG4}, but  will  focus on the measurement of the momentum shift operator $\hat U$ from the main text. We will follow the notation and graphical representation of \cite{LCRG2,LCRG3, LCRG4}.	
\\[1ex]
%%%%%%%%%%%%%%%%%%%%%%%%%%%%%%%%%%%%%%%%%%%%%%%%%
%%%%%%%%%%%%%%%%%%%%%%%%%%%%%%%%%%%%%%%%%%%%%%%%%

\begin{figure}[h]
	\includegraphics[width=0.6\textwidth]{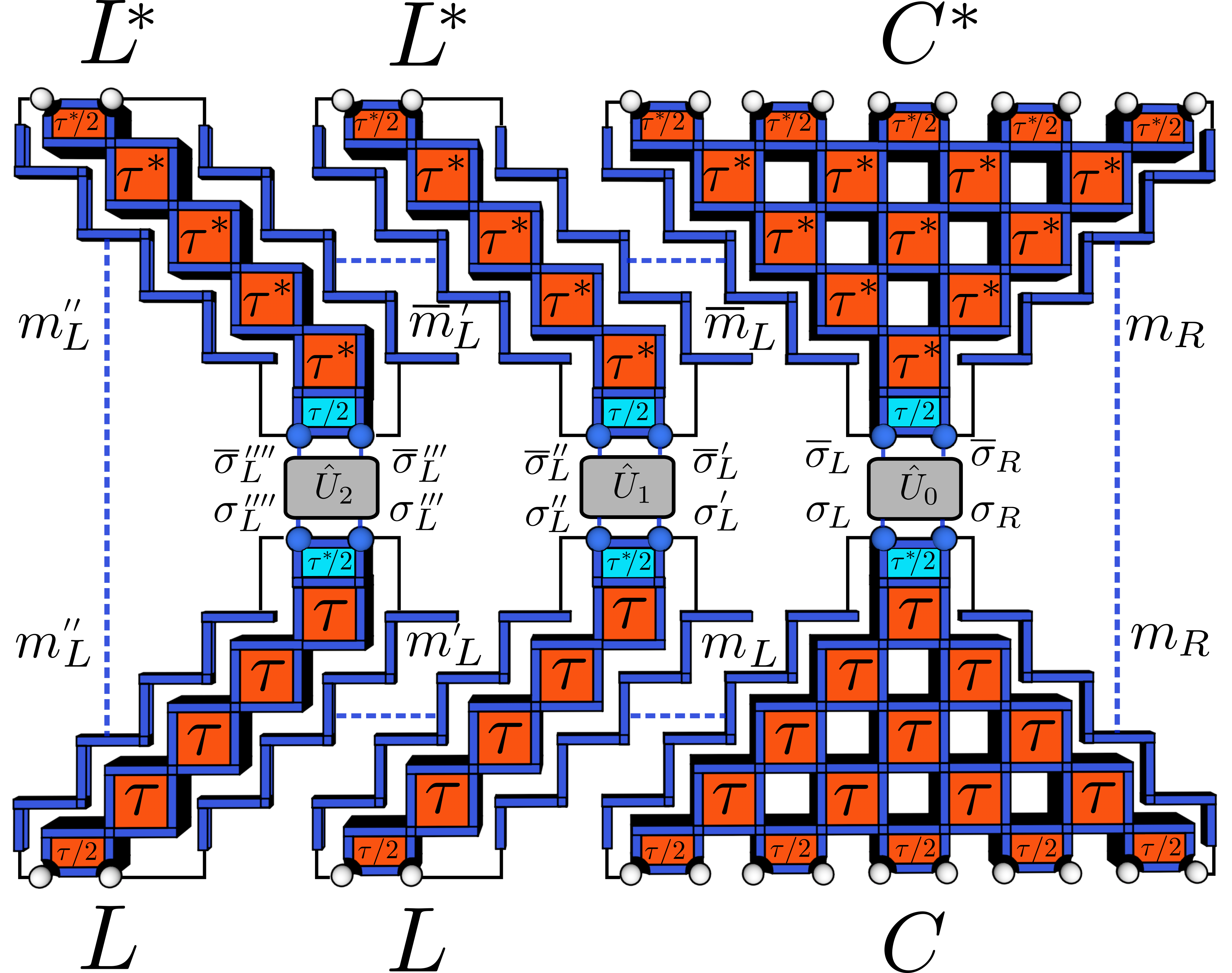}
	\caption{LCRG tensor-object to measure the momentum shift operator.  Blue spheres describe local directly accessible bonds, containing a unit cell or singular sites, e.g. $\sigma_L,\sigma_R,$. All other bonds are truncated and included in the blue superblocks. The white spheres are a guide to the eye representing the initial state and the actual size of the system (white spheres amount of sites or unit cells).
		The time-evolution is done by Trotter-Suzuki plaquettes, where red plaquettes evolve $\textbf{C}$ forward in time and blue plaquettes backwards in time. This has the advantage of using a second-order trotter decomposition at the speed of a first order decomposition \cite{LCRG4}.
		Here it is shown how to contract the tensor objects to measure the operator $\hat U$ with the tensor objects $\textbf{L}$ and $\textbf{C}$. The left transfer matrices $\textbf{L}^*(t),\textbf{L}(t)$ are applied twice to the local density operator created from $\textbf{C}^*\textbf{C} (t)$, see eq. \ref{local}, encapsuling $\hat U$. Parts of this picture have been taken from \cite{LCRG4} and have been modified.  }
	\label{corrfunc}
\end{figure}

The LCRG has three major components. The two transfer matrices for enlarging the system to the left $ \textbf{L}$ and right $ \textbf{R}$ and the actual state in a light cone shape named $ \textbf{C}$. All components have 4 indices, which are explained in the following. From a numerical viewpoint, the components are therefore refered to as tensor objects of the LCRG. The time-evolution is  done by Trotter-Suzuki plaquettes applied to the initial tensor objects \cite{LCRG2,LCRG3}, see fig. \ref{corrfunc}, (red and blue plaquettes).

The standard tensor then has two local bonds, denoted by $\sigma_L$, $\sigma_R$ in the index and two superblocks, denoted by $m_L$, $m_R$ in the index. Blue dotted lines in Fig.\ref{corrfunc} imply the summation over pairwise indices. In the numerics of the main text one bond contains two sites, as well as two additional sites for the auxilliary system needed for the purification ansatz. Thus a bond has dimension $d=2^4$, since we are only considering hardcore bosons in the numerics. We keep around 300 states for the superblocks $m_L$ and $m_R$, where we have checked convergence and made sure not to exceed a discarded weigth of $10^{-10}$ for the first period of the Thouless pump and $10^{-8}$ for the second. % All further details of the algorithm can be found in \cite{LCRG2,LCRG3, LCRG4}.
%	In the following we want to clarify how we have computed the momentum shift operator, also shown in eq. (1) of the main text and from this the EGP $\phi_\textrm{EGP} = 2\pi P = \textrm{Im} \log \textrm{Tr}\bigl\{ \rho \, \hat U\bigr\}$. 
The momentum shift operator, needed to calculate the EGP of the main text, reads
\begin{equation}
\hat U = e^{ i 2\pi \hat X},\quad \textrm{with}\quad\hat X = \frac{1}{L}\sum_{j=1}^L\sum_{s=1}^n \left(j + r_s\right) \hat n_{js},
\end{equation}
where $\hat n_{js}$ denotes the particle number operator of the $s$th site $(s\in\{1,2,\dots,n\})$ in the $j$th unit cell. $0\le r_s \le 1$ characterizes the position within the unit cell. Here we set the lattice constant to 1. When using the purification ansatz to account for finite temperature, we need to measure $\hat n_{js} \otimes \mathbb{I}$, where $\mathbb{I}$ is the identity matrix applied to the auxiliary system effectively tracing out the ancillary sites. 
Since the operator $\hat n_{js}$ is diagonal in real space, we can rewrite $\hat U$ as a product 
\begin{equation}
\hat U = \prod _{j=1} ^ L \hat U _j = \prod _{j=1} ^ L e^{ i 2\pi \hat X_j},\quad \textrm{with}\quad\hat X_j = \frac{1}{L}\sum_{s=1}^n \left(j + r_s\right) \hat n_{js}.
\end{equation}
In the LCRG we can compute $\hat U$ e.g. by multiplying left transfer matrices to extend the local density matrix.
\begin{align}
\rho_{\mathrm{local}}[\bar m_L,\bar \sigma_{L},\bar \sigma_{R},m_L,\sigma_{L},\sigma_{R}]  = \sum_{m_R} C^*[\bar m_L,\bar \sigma_{L},\bar \sigma_{R},m_R]\  C[m_L,\sigma_{L},\sigma_{R},m_R]
\label{local}
\end{align}
\begin{align}
\begin{split}
(\rho U_0)_\mathrm{extend}[\bar m_L', \bar \sigma_L'', \bar \sigma_L', m_L', \sigma_L'', \sigma_L' ] = \sum_{\bar m_L,m_L,\bar \sigma_{L},\bar \sigma_{R}, \sigma_{R},\sigma_{L}}& L^*[\bar m_L', \bar \sigma_L'', \bar \sigma_L',\bar m_L]  L[ m_L',  \sigma_L'',  \sigma_L', m_L] \times \\&\times \rho_{\mathrm{local}}[\bar m_L,\bar \sigma_{L},\bar \sigma_{R},m_L,\sigma_{L},\sigma_{R}]  U_0 [\bar \sigma_{L},\bar \sigma_{R}, \sigma_{L},\sigma_{R}].
\end{split}
\label{enlargen}
\end{align}
Applying eq. (\ref{enlargen}) twice, changing $U_0$ to $U_1, U_2$ etc. and then tracing out the most left superblock, will leave us with the tensor object presented in fig. \ref{corrfunc}. By applying eq. (\ref{enlargen}) more than twice we can receive any wanted block size. 

%%%%%%%%%%%%%%%%%%%%%%%%%
%%%%%%%%%%%%%%%%%%%%%%%%%%%%%%%%%%%%%%%

%%%%%%%%%%%%%%%%%%%%%%%%%%%%%%%%%%%%%%%
\section{Flattened Hamiltonian}
%%%%%%%%%%%%%%%%%%%%%%%%%%%%%%%%%%%%%%%

Here we proof that $\vert \textrm{Tr}\{(\hat U)^d\}\vert\sim {\cal O}(d)$, as argued in the main text. For this we consider a system of $N$ particles in a lattice with $L$ unit cells
and a fractional filling of $1/d$ per unit cell, i.e. $N= L/d$. 
\begin{eqnarray*}
	\textrm{Tr}\{(\hat U)^d\} = \textrm{Tr}\left\{\exp\left(\frac{2\pi i}{L} d \sum_{j=1}^L\sum_{s} (j + r_s) \hat n_{js}\right)\right\}.
\end{eqnarray*}
The constraint of constant particle number can most conveniently be written in terms of an integral of an angle $\theta$:
\begin{eqnarray*}
	\textrm{Tr}\{(\hat U)^d\} =  \frac{1}{2\pi}\int_0^{2\pi}\!\!\! d\theta\, e^{iN\theta} \prod_{j,s} \sum_{n} 
	e^{-i n(\theta -\phi_s)}
	\exp\left\{\frac{2\pi i}{N} n\, j\right\}
\end{eqnarray*}
where $\phi_s = 2\pi r_s/N$. In the case of fermions the sums over occupation numbers $n$ extends over $0$ and $1$ and we find
\begin{eqnarray*}
	&&\textrm{Tr}\{(\hat U)^d\} =\frac{1}{2\pi}\int_0^{2\pi}\!\!\! d\theta\, e^{iN\theta} \prod_{s} 
	\left[\prod_{j=1}^L\left(1+e^{-i(\theta-\phi_s)} e^{2\pi i j/N}\right)\right]\\
	&&\quad= \frac{1}{2\pi}\int_0^{2\pi}\!\!\! d\theta\, e^{iN\theta} \prod_{s} 
	\left[\prod_{j=1}^N\left(1+e^{-i(\theta-\phi_s)} e^{2\pi i j/N}\right)\right]^d
\end{eqnarray*}
By introducing $z=-e^{-i\left(  \theta-\phi_{s}\right)  }$ the product incide the bracket can be
simplified to
\[%
%TCIMACRO{\dprod \limits_{j=1}^{N}}%
%BeginExpansion
{\displaystyle\prod\limits_{j=1}^{N}}
%EndExpansion
\left(  1+e^{-i\left(  \theta-\phi_{s}\right)  }e^{2\pi ij/N}\right)  =%
%TCIMACRO{\dprod \limits_{j=1}^{N}}%
%BeginExpansion
{\displaystyle\prod\limits_{j=1}^{N}}
%EndExpansion
\left(  1-ze^{2\pi ij/N}\right)  =
\]%
\[%
%TCIMACRO{\dprod \limits_{j=1}^{N}}%
%BeginExpansion
{\displaystyle\prod\limits_{j=1}^{N}}
%EndExpansion
\left(  e^{-2\pi ij/N}-z\right)  =1-z^{N}.
\]
We have used the fact that $e^{-2\pi ij/N}$ are complex roots of $1-z^{N}=0$. Thus we find
\begin{eqnarray*}
	\textrm{Tr}\{(\hat U)^d\} =\frac{1}{2\pi}\int_0^{2\pi}\!\!\! d\theta\, e^{iN\theta} \prod_{s} \Bigl[1-(-1)^N e^{-i\theta N} e^{i\phi_s N}\Bigr]^d
\end{eqnarray*}
The only non-vanishing contribution to the  $\theta$ integration are terms  in the product that contain $e^{-i\theta N}$ to the first power.
Thus we find
\begin{equation}
\vert \textrm{Tr}\{(\hat U)^d\} \vert = d \sum_{s} e^{2\pi i r_s} = {\cal O}(d).
\end{equation}
We note that for special cases the sum over the sites within a unit cell can vanish, but generically $\vert \textrm{Tr}\{(\hat U)^d\} \vert $ is on the oder
of $d$. A similar calculation can be made for bosons where the sums over occupation numbers runs from $0$ to $N$.

%%%%%%%%%%%%%%%%%%%%%%%%%%%%%%%%%%%%%%%%%%%%%%%
\section{Scaling of $\vert z(T,L)\vert$  for non-interacting fermions}
%%%%%%%%%%%%%%%%%%%%%%%%%%%%%%%%%%%%%%%%%%%%%%%

Here we discuss the scaling of $\vert z(T,L)\vert$ for the case of a two-band model of non-interacting fermions at half filling.	
($N=L,$ where $N$ is the total number of fermions and $L$ is
the number of unit cells in the system)
As shown in Ref.[17] of the main paper, the many-body polarization 
can be represented in the following form
\begin{equation}
z=\frac{\left\langle R\right\rangle }{\mathcal{Z}},\label{Polarization}%
\end{equation}
where
\begin{equation}
\left\langle R\right\rangle =\mathtt{tr}\left[
{\displaystyle\prod\limits_{k=1}^{N}}
\exp\left(  \beta\overrightarrow{\alpha}_{k}\overrightarrow{\sigma}\right)
\right]  ,\label{modul_trace}%
\end{equation}
and $\mathcal{Z}$ is the parition function%
\begin{equation}
\mathcal{Z}=\mathtt{tr}\exp\left(  -\beta H\right)  =
{\displaystyle\prod\limits_{k=1}^{N}}
\left(  {{1}}+e^{\beta\left\vert \overrightarrow{\alpha}_{k}\right\vert
}\right)
{\displaystyle\prod\limits_{k=1}^{N}}
\left(  {{1}}+e^{-\beta\left\vert \overrightarrow{\alpha}_{k}\right\vert
}\right)  .\label{Partition_function2}
\end{equation}
The Hamiltonian $H$ can be represented as a direct sum of the the
momentum-space Hamiltonian matrix $h_{k}=\overrightarrow{\alpha}_{k}\cdot\overrightarrow{\sigma}$,
where $\overrightarrow{\sigma}$ is the vector of Pauli matrices and $\overrightarrow{\alpha}_{k}$
are real vectors.

Our ultimate goal is to investigate the amplitude of $z$ in the thermodynamic
limit i.e. when $N=\frac{M}{2}>>1$. Following Ref.[17]  of the main paper, 
expression (\ref{modul_trace}) can be represented as
\begin{equation}
\left\langle R\right\rangle =\mathtt{tr}\left[
{\displaystyle\prod\limits_{k=1}^{N}}
\exp\left(  \beta\left\vert \overrightarrow{\alpha}_{k}\right\vert \sigma
_{z}\right)  \cdot u_{k-1}u_{k}^{\dagger}\right]  ,\label{unitary}%
\end{equation}
where the unitary matrix $u_{k}$ transforms the Bloch Hamiltonian
$h_k$ into a diagonal matrix i.e.$
\overrightarrow{\alpha}_{k}\overrightarrow{\sigma}=\left\vert \overrightarrow
{\alpha}_{k}\right\vert u_{k}^{\dagger}\sigma_{z}u_{k}$.
The trace in eq.(\ref{unitary}) can be calculated easily in the thermodynamic limit
i.e. $N\rightarrow\infty$. Indeed,
\begin{equation}
u_{k-1}u_{k}^{\dagger}\approx1+\frac{2\pi}{N}u_{k-1}\frac{\partial
	u_{k}^{\dagger}}{\partial k}+...\underset{N>>1}{\rightarrow}%
1.\label{limit_large_N}%
\end{equation}
Using this asymptotic form the module of $\left\langle R\right\rangle $ takes
the following simple form%
\begin{eqnarray*}
	\left\vert \left\langle R\right\rangle \right\vert \approx\exp\left(
	{\displaystyle\sum\limits_{k=1}^{N}}
	\beta\left\vert \overrightarrow{\alpha}_{k}\right\vert \right)  +\exp\left(  -{\displaystyle\sum\limits_{k=1}^{N}}
	\beta\left\vert \overrightarrow{\alpha}_{k}\right\vert \right)  +\mathcal{O}%
	\left(  \frac{1}{N}\right)  .
\end{eqnarray*}
Then, using Eq. (\ref{Partition_function2}), one has
\begin{equation}
\left\vert z\right\vert \approx\frac{\exp\left(
	{\displaystyle\sum\limits_{k}}
	\beta\left\vert \overrightarrow{\alpha}_{k}\right\vert \right)  +\exp\left(  -%
	{\displaystyle\sum\limits_{k}}
	\beta\left\vert \overrightarrow{\alpha}_{k}\right\vert \right)  }{%
	{\displaystyle\prod\limits_{k}}
	\left(  2+e^{\beta\left\vert \overrightarrow{\alpha}_{k}\right\vert
	}+e^{-\beta\left\vert \overrightarrow{\alpha}_{k}\right\vert }\right)
}.\label{zet_amplitude}%
\end{equation}
We proceed by giving estimates for the lower and upper bounds of
$\left\vert z\right\vert $.

\

\paragraph{Lower bound: }

\begin{eqnarray}
&&\left\vert z\right\vert =\frac{1+\exp\left(  -2\beta
	{\sum\limits_{k=1}}
	\left\vert \overrightarrow{\alpha}_{k}\right\vert \right)  }{%
	{\prod\limits_{k}}
	\left(  2e^{-\beta\left\vert \overrightarrow{\alpha}_{k}\right\vert
	}+1+e^{-2\beta\left\vert \overrightarrow{\alpha}_{k}\right\vert }\right)
}\geq\frac{1}{
	{\prod\limits_{k}}
	\left(  2e^{-\beta\left\vert \overrightarrow{\alpha}_{k}\right\vert
	}+1+e^{-2\beta\left\vert \overrightarrow{\alpha}_{k}\right\vert }\right)
}\geq \nonumber\\
&&\frac{1}{\left(  1+\frac{2
		\sum\limits_{k=1}
		e^{-\beta\left\vert \overrightarrow{\alpha}_{k}\right\vert }+%
		\sum\limits_{k=1}
		e^{-2\beta\left\vert \overrightarrow{\alpha}_{k}\right\vert }}{N}\right)
	^{N}}\underset{N_{F}>>1}{\rightarrow}\frac{1}{\exp\left(  2%
	\sum\limits_{k}
	e^{-\beta\left\vert \overrightarrow{\alpha}_{k}\right\vert }+%
	\sum\limits_{k}
	e^{-2\beta\left\vert \overrightarrow{\alpha}_{k}\right\vert }\right)
},\label{Lower_Bound}%
\end{eqnarray}
where in the first line we have used the arithmetic-geometric inequality.

\

\textit{Upper bound: }To derive an upper bound on $\left\vert z\right\vert $,
we note first that the function $\ln\left(  2e^{-x}+1+e^{-2x}\right)  $ is a
convex function and by Jensen's inequality
\begin{eqnarray}
&&\ln\left(
{\prod\limits_{k}}
\left(  2e^{-\beta\left\vert \overrightarrow{\alpha}_{k}\right\vert
}+1+e^{-2\beta\left\vert \overrightarrow{\alpha}_{k}\right\vert }\right)
\right)  =\nonumber\\
&&=
{\sum\limits_{k}}
\ln\left(  2e^{-\beta\left\vert \overrightarrow{\alpha}_{k}\right\vert
}+1+e^{-2\beta\left\vert \overrightarrow{\alpha}_{k}\right\vert }\right)
\geq\ln\left(  1+e^{-\beta\Delta}\right)  ^{2N},\label{upper_bound}%
\end{eqnarray}
where $\Delta=\frac{1}{N}{\sum\limits_{k}}
\left\vert \overrightarrow{\alpha}_{k}\right\vert$
is the spectral gap of the Hamiltonian. We thus have the following upper bound
\begin{equation}
\left\vert z\right\vert \leq\frac{1+\exp\left(  -2N\beta\Delta\right)
}{\left(  1+e^{-\beta\Delta}\right)  ^{2N}}.\label{upper_bound33}%
\end{equation}
Therefore, 
\begin{equation}
\exp\left(  -
{\sum\limits_{k}}
e^{-2\beta\left\vert \overrightarrow{\alpha}_{k}\right\vert }-2{\sum\limits_{k}}
e^{-\beta\left\vert \overrightarrow{\alpha}_{k}\right\vert }\right)
\leq\left\vert z\right\vert \leq\frac{1+\exp\left(  -2N\beta\Delta\right)
}{\left(  1+e^{-\beta\Delta}\right)  ^{2N}}.\label{Bounds33}
\end{equation}
We verify immediately that if
\begin{equation}
{\sum\limits_{k}}
e^{-\beta\left\vert \overrightarrow{\alpha}_{k}\right\vert }<\infty
,\label{condition}%
\end{equation}
the absolute value of $z$ never vanishes, i.e. $\left\vert z\right\vert >0$ even when $N\rightarrow
\infty$. However, in general 
condition (\ref{condition}) is violated and there might be an exact zero of $\vert z\vert$.

Taking $\left\vert \overrightarrow{\alpha}_{k}\right\vert $outside the
summation in the above inequality corresponds to assuming a flat single particle spectrum.
In the low temperature limit, $\beta\Delta>>1$, this is well justified and it is easily seen that upper and lower
bounds of $\left\vert z\right\vert $ are the same giving the general scaling
\begin{equation}
\left\vert z\right\vert \approx\exp\left(  -2Ne^{-\beta\Delta}\right)
.\label{estimation}%
\end{equation}
Hence, despite of the fact that $\left\vert z\right\vert $ is exponentially
small in system size, the relevant temperature, where $\vert z(T)\vert$ deviates from unity scales only logarithmically
in system size  i.e. $T\sim\Delta/\ln\left(  2N\right)  $.

\twocolumngrid

\end{document}